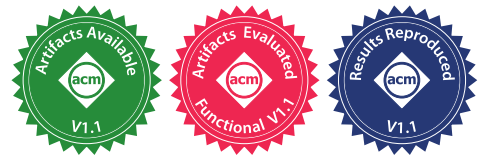

# CutQC: Using Small Quantum Computers for Large Quantum Circuit Evaluations

Wei Tang
weit@princeton.edu
Department of Computer Science,
Princeton University
Mathematics and Computer Science
Division, Argonne National
Laboratory
USA

Teague Tomesh
ttomesh@princeton.edu
Department of Computer Science,
Princeton University
Mathematics and Computer Science
Division, Argonne National
Laboratory
USA

Martin Suchara
msuchara@anl.gov
Mathematics and Computer Science
Division, Argonne National
Laboratory
USA

Jeffrey Larson
jmlarson@anl.gov
Mathematics and Computer Science
Division, Argonne National
Laboratory
USA

Margaret Martonosi
mrm@princeton.edu
Department of Computer Science,
Princeton University
USA

## ABSTRACT

Quantum computing (QC) is a new paradigm offering the potential of exponential speedups over classical computing for certain computational problems. Each additional qubit doubles the size of the computational state space available to a QC algorithm. This exponential scaling underlies QC's power, but today's Noisy Intermediate-Scale Quantum (NISQ) devices face significant engineering challenges in scalability. The set of quantum circuits that can be reliably run on NISQ devices is limited by their noisy operations and low qubit counts.

This paper introduces CutQC, a scalable hybrid computing approach that combines classical computers and quantum computers to enable evaluation of quantum circuits that cannot be run on classical or quantum computers alone. CutQC cuts large quantum circuits into smaller subcircuits, allowing them to be executed on smaller quantum devices. Classical postprocessing can then reconstruct the output of the original circuit. This approach offers significant runtime speedup compared with the only viable current alternative—purely classical simulations—and demonstrates evaluation of quantum circuits that are larger than the limit of QC or classical simulation. Furthermore, in real-system runs, CutQC achieves much higher quantum circuit evaluation fidelity using small prototype quantum computers than the state-of-the-art large NISQ devices achieve. Overall, this hybrid approach allows users to leverage classical and quantum computing resources to evaluate quantum programs far beyond the reach of either one alone.

**ACM Reference Format:**
Wei Tang, Teague Tomesh, Martin Suchara, Jeffrey Larson, and Margaret Martonosi. 2021. CutQC: Using Small Quantum Computers for Large Quantum Circuit Evaluations. In *Proceedings of the 26th ACM International Conference on Architectural Support for Programming Languages and Operating Systems (ASPLOS '21), April 19–23, 2021, Virtual, USA.* ACM, New York, NY, USA, 14 pages. https://doi.org/10.1145/3445814.3446758

## 1 INTRODUCTION

Quantum computing (QC) has emerged as a promising computational approach with the potential to benefit numerous scientific fields [41]. For example, QC offers the possibility of reduced computational time for problems in machine learning [8, 27], chemistry [1, 25], and other areas [31]. Some of the earliest QC work shows that quantum algorithms for factoring [43] can be exponentially faster and that database search [17] can be polynomially faster than their classical counterparts. However, these quantum algorithms assume the existence of large-scale, fault-tolerant, universal quantum computers.

Instead, today's quantum computers are noisy intermediate-scale quantum (NISQ) devices [39], and major challenges limit their effectiveness. Noise can come from limited coherence time [24], frequency selection for individual qubits [26], crosstalk among qubits [33], and limited control bandwidth [42]. Because of these and other issues, the difficulty of building reliable quantum computers increases dramatically with increasing number of qubits. For example, Figure 1 shows the fidelities obtained from executions of the Bernstein–Vazirani (BV) algorithm on IBM quantum computers with increasing number of qubits. We executed quantum circuits of only half the size of the devices themselves and mapped the

Wei Tang, Teague Tomesh, Martin Suchara, Jeffrey Larson, and Margaret Martonosi

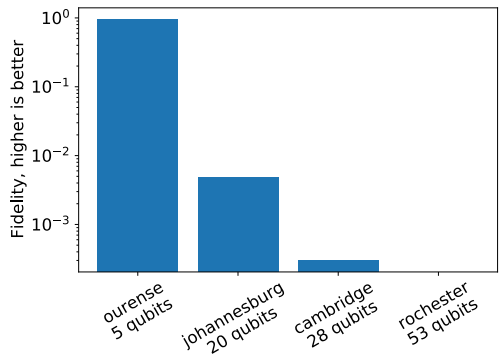

**Figure 1: Using IBM's quantum computers to execute the Bernstein–Vazirani (BV) algorithm. Problem instance sizes were selected to occupy half of the device qubits. Fidelity (correct answer probability) decreases rapidly for larger devices and drops below 1% for a 10-qubit BV executed on the 20-qubit Johannesburg device. The 53-qubit Rochester device fails to produce any meaningful results for a 26-qubit BV circuit.**

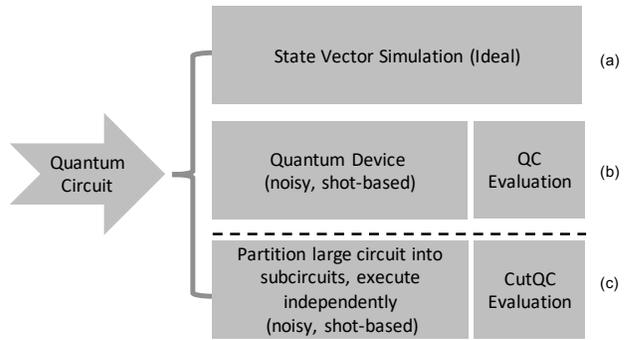

**Figure 2: Different quantum circuit evaluation modes. (a) Purely classical simulation produces the ground truth to verify other evaluation outputs. (b) Purely quantum evaluation on quantum computers. Multiple vendors provide cloud access to their devices. (c) Our hybrid mode, which is orders of magnitude faster than (a), produces much less noisy outputs than (b), and evaluates much larger circuits than (a) and (b).**

circuits to use the best qubits on the devices by using a state-of-the-art noise-adaptive compiler [32]. However, larger devices realize significantly worse fidelity than do smaller devices.

More fundamentally, such intermediate-scale quantum devices are hard limited by their qubit count. Currently, only small quantum circuits can be run on small quantum computers. The largest superconducting quantum computers available today have 53 qubits [3, 21], and their relatively poor fidelity further limits the size of circuits that can be run. Large neutral atom qubit arrays have been developed recently, but achieving high gate fidelity remains a significant challenge [16].

Both the noise and the intermediate-scale characteristics of NISQ devices present significant obstacles to their practical applications. On the other hand, the only currently viable alternative for QC evaluation—classical simulations of quantum circuits—produces noiseless output but is not tractable in general. For example, state-of-the-art full-state classical simulations of quantum circuits of merely 45 qubits require tens of hours on thousands of high-performance compute nodes and hundreds of terabytes of memory [52].

This work uses circuit cutting to expand the reach of small quantum computers with partitioning and postprocessing techniques that augment small QC platforms with classical computing resources. We develop the first end-to-end hybrid approach that automatically locates efficient cut positions to cut a large quantum circuit into smaller subcircuits that are each independently executed by using quantum devices with fewer qubits. Via scalable postprocessing techniques, the output of the original circuit can then be reconstructed or sampled efficiently from the subcircuit outputs.

To evaluate the performance of CutQC, we benchmarked six different quantum circuits that represent a general set of circuits for gate-based QC platforms and promising near-term applications. We demonstrate executing quantum circuits of up to 100 qubits on existing NISQ devices. This is significantly beyond the current reach of either quantum or classical methods alone.

Our contributions include the following:

(1) **Expanding the size** of quantum circuits that can be run on NISQ devices and classical simulation by combining the two. Our method allows executions of quantum circuits more than twice the size of the available quantum computer backend and much beyond the classical simulation limit.

(2) **Improving the fidelity** of quantum circuit executions on NISQ devices. We show an average of 21% to 47% improvement to $\chi^2$ loss for different benchmarks by using CutQC with small quantum computers, as compared with direct executions on large quantum computers.

(3) **Speeding up** the overall quantum circuit execution over classical simulations. We use quantum computers as coprocessors to achieve 60X to 8600X runtime speedup over classical simulations for different benchmarks.

## 2 BACKGROUND

This section introduces quantum circuits and explains the differences between several quantum circuit evaluation modes. For a more comprehensive introduction to quantum computing we refer the reader to [13, 34].

Quantum programs are expressed as circuits that consist of a sequence of single- and multiqubit gate operations. Quantum circuits can be evaluated by using classical simulations, on quantum computers, or in a hybrid mode as explored in this paper. Figure 2 provides an overview of the different evaluation modes. Several simulation and hardware QC platforms recently emerged [15, 29, 40]. One widely used package is the IBM Qiskit [2], which allows simulation and cloud access to IBM's quantum hardware.

State vector simulation (Figure 2a) is typically an idealized noiseless simulation of a quantum circuit. All quantum operations are represented as unitary matrices. N-qubit operations are $2^N \times 2^N$ unitary matrices. State vector simulation executes circuits by sequentially multiplying each gate's corresponding unitary matrix with the current state vector. This yields an error-free output represented as complex amplitudes, which cannot be obtained on quantum computers. This evaluation mode scales exponentially and serves only to provide the ground truth for benchmarking NISQ devices for small quantum circuits. We use this evaluation mode as a baseline to verify the output of modes (b) and (c) in Figure 2 and to compute the $\chi^2$ metric to quantify the noise and quality of quantum circuit executions.

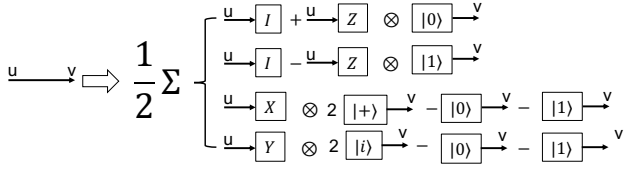

**Figure 3: Procedure to cut one qubit wire. The wire between vertices $u$ and $v$ (left) can be cut by (as shown on the right) summing over four pairs of measurement circuits appended to $u$ and state initialization circuits prepended to $v$. Measurement circuits in the $I$ and $Z$ basis have the same physical implementation. The three different upstream measurement circuits and the four different downstream initialization circuits are now separate and can be independently evaluated.**

Physical executions on NISQ computers use a shot-based model. Quantum algorithms are first compiled to satisfy device-specific characteristics such as qubit connectivity, native gate set, noise, and crosstalk [32, 33]. A real NISQ device then executes the compiled quantum circuit thousands of times ("shots") in quick succession. At the end of each shot, all qubits are measured; and the output, a classical bit string, is recorded. After all shots are taken, a distribution of probabilities over the observed states is obtained. Section 6 compares the runtimes of the state vector simulation (Figure 2a) and CutQC evaluation (Figure 2c) modes. We also compare the execution fidelities of the QC evaluation (Figure 2b) and CutQC evaluation (Figure 2c) modes.

## 3 CIRCUIT CUTTING

This section presents an overview of the theory behind cutting a quantum circuit. Figure 4 offers an illustrative example, where one cut separates a 5-qubit quantum circuit into 2 subcircuits of 3 qubits each. Time goes from left to right in quantum circuit diagrams, and each row represents a qubit wire, in other words, timewise cuts. CutQC performs vertical cuts on qubit wires, in other words, timewise cuts.

### 3.1 Circuit Cutting: Theory

The physics theory behind the ability to cut a qubit wire originates from the fact that the unitary matrix of an arbitrary quantum operation in a quantum circuit can be decomposed into any set of orthonormal matrix bases. For example, the set of Pauli matrices $I, X, Y, Z$ is a convenient basis to use. Previous work in theoretical physics proved the mathematical validity of decomposing unitary matrices of quantum operations but with an exponentially higher overhead [37].

Specifically, an arbitrary 2×2 matrix $\mathbf{A}$ can be decomposed as

$$\mathbf{A} = \frac{Tr(\mathbf{A}I)I + Tr(\mathbf{A}X)X + Tr(\mathbf{A}Y)Y + Tr(\mathbf{A}Z)Z}{2}. \quad (1)$$

This identity, however, requires access to complex amplitudes, which are not available on quantum computers. To execute on quantum computers, we further decompose the Pauli matrices into their eigenbasis and organize the terms. We obtain the following identity in cutting a quantum wire timewise.

$$\mathbf{A} = \frac{A_1 + A_2 + A_3 + A_4}{2} \quad (2)$$

where

$$A_1 = [Tr(\mathbf{A}I) + Tr(\mathbf{A}Z)]|0\rangle\langle0|$$
$$A_2 = [Tr(\mathbf{A}I) - Tr(\mathbf{A}Z)]|1\rangle\langle1|$$
$$A_3 = Tr(\mathbf{A}X)[2|+\rangle\langle+| - |0\rangle\langle0| - |1\rangle\langle1|]$$
$$A_4 = Tr(\mathbf{A}Y)[2|+i\rangle\langle+i| - |0\rangle\langle0| - |1\rangle\langle1|]$$

Each trace operator corresponds physically to measure the qubit in one of the Pauli bases. And each of the density matrices corresponds physically to initialize the qubit in one of the eigenstates. Figure 3 shows the resulting subcircuits and the reconstruction procedure incurred when making a single cut. Since measuring a qubit in either the $I$ or $Z$ basis corresponds physically to the same quantum circuit, three different upstream subcircuits and four different downstream subcircuits result. Four pairs of Kronecker products between the subcircuit outputs are then performed and summed together to reconstruct the uncut circuit output. A similar procedure can then be applied to more than one cutting point in a large quantum circuit in order to split it into a few smaller subcircuits.

### 3.2 Circuit Cutting: Example

Consider the quantum circuit example in Figure 4. Here we show how the example 5-qubit circuit can be cut to fit on a 3-qubit device. First, we define notation for a circuit's output state probability distribution. Let the input to an $n$-qubit circuit be initialized to the $|q_0, \ldots, q_{n-1}\rangle$ state, where $q_i \in \{|0\rangle, |1\rangle, |+\rangle, |+i\rangle\}$. Let the output be measured in the $M_0, \ldots, M_{n-1}$ basis, where $M_i \in \{I, X, Y, Z\}$. We use the notation $C(|q_0, \ldots, q_{n-1}\rangle; M_0, \ldots, M_{n-1})$ to represent a quantum circuit $C$ with its qubits initialized in the given states and measured in the given basis.

*3.2.1 Selecting the Cut Locations.* Assuming for now that cut locations are chosen manually, we show in Figure 4 that a single cut can be made to qubit $q_2$ between the first two $cZ$ gates, splitting the original 5-qubit circuit into two circuits containing 3 qubits each. Now, the last qubit in subcircuit 1 ($subcirc1_2$) and the first qubit in subcircuit 2 ($subcirc2_0$) can be mapped to the $u$ and $v$ appearing in the right-hand side of Figure 3. Section 4.1 describes the automation of the selection of cut locations.

*3.2.2 Attributing the Shots.* Note that $subcirc1_2$ does not appear in the final output of the uncut circuit. Therefore each shot obtained from executing the subcircuit 1 needs to be multiplied by a ±1 factor, contingent on the measurement outcomes of qubit $subcirc1_2$. Specifically, each measurement outcome of subcircuit 1 should be

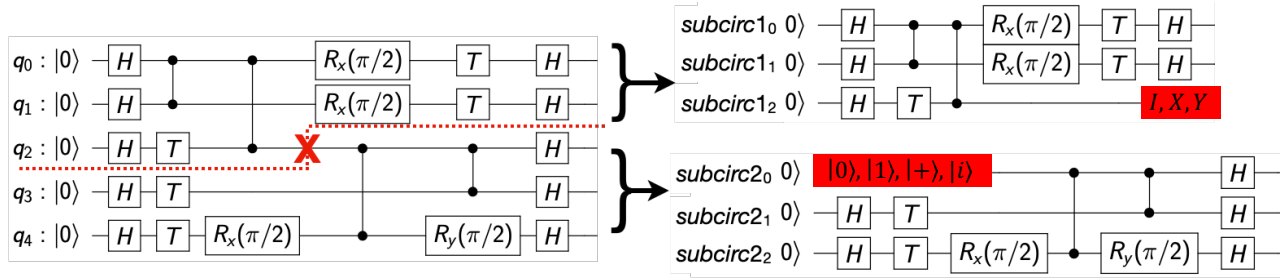

**Figure 4: Example of cutting a five-qubit circuit into two smaller subcircuits of three qubits each. The subcircuits are produced by cutting the $q_2$ wire between the first two $cZ$ gates. The three variations of $subcircuit_1$ and four variations of $subcircuit_2$ can then be evaluated on a 3-qubit quantum device, instead of a 5-qubit device. The classical postprocessing involves summing over four Kronecker products between the two subcircuits for the one cut made.**

attributed to the final output as

$$
\begin{cases}
\overline{xx0}, \overline{xx1} \rightarrow +\overline{xx} & M_2 = I \\
\overline{xx0} \rightarrow +\overline{xx} \\
\overline{xx1} \rightarrow -\overline{xx} & \text{otherwise,}
\end{cases}
\tag{3}
$$

where $\overline{xx}$ is the measurement outcome of the qubits $subcirc1_0$ and $subcirc1_1$.

We demonstrate an example of how the probability of the state $|01010\rangle$ of the uncut circuit is calculated. The relevant state of subcircuit 1 is $|01\rangle$. According to Equation 3, the four subcircuit 1 terms involved in the reconstruction are

$$
\begin{aligned}
p_{1,1} &= p(|010\rangle |I\rangle) + p(|011\rangle |I\rangle) + p(|010\rangle |Z\rangle) - p(|011\rangle |Z\rangle) \\
p_{1,2} &= p(|010\rangle |I\rangle) + p(|011\rangle |I\rangle) - p(|010\rangle |Z\rangle) + p(|011\rangle |Z\rangle) \\
p_{1,3} &= p(|010\rangle |X\rangle) - p(|011\rangle |X\rangle) \\
p_{1,4} &= p(|010\rangle |Y\rangle) - p(|011\rangle |Y\rangle).
\end{aligned}
$$

The relevant state of subcircuit 2 is $|010\rangle$. Hence, its four terms are

$$
\begin{aligned}
p_{2,1} &= p(|010\rangle \,|\, |0\rangle) \\
p_{2,2} &= p(|010\rangle \,|\, |1\rangle) \\
p_{2,3} &= 2p(|010\rangle \,|\, |+\rangle) - p(|010\rangle \,|\, |0\rangle) - p(|010\rangle \,|\, |1\rangle) \\
p_{2,4} &= 2p(|010\rangle \,|\, |i\rangle) - p(|010\rangle \,|\, |0\rangle) - p(|010\rangle \,|\, |1\rangle).
\end{aligned}
$$

*3.2.3 Building the Full Probabilities.* The full probability distribution for the uncut circuit can then be reconstructed in the classical postprocessing step by taking the relevant outputs of the two smaller subcircuits, performing the four pairs of Kronecker products, and summing together, as indicated in Figure 3. Specifically, the final reconstructed probability of the uncut state $|01010\rangle$ is

$$
p(|01010\rangle) = \frac{1}{2} \sum_{i=1}^{4} p_{1,i} \, p_{2,i}.
$$

The mathematical theory of circuit cutting [37] proves that the CutQC output strictly equals the output of the uncut circuit. However, if too few shots were taken for the subcircuits, the subcircuit probabilities can be far from convergence. As a result, it is possible to get a negative reconstructed probability output. However, much like a user is expected to take enough shots when evaluating an uncut circuit, one is also expected to take sufficient shots for the subcircuits in the CutQC mode. Our real-device experiments took at most 8,192 shots for one subcircuit, depending on the size of the subcircuit, and we did not observe negative results.

### 3.3 Circuit Cutting: Challenges

The first challenge is to find cut locations. While quantum circuits can always be split into smaller ones, finding optimal cut locations is crucial in order to minimize the classical postprocessing overhead. In general, large quantum circuits may require more than one cut in order to be separated into subcircuits. In this case, the cutting scheme evaluates all possible measurement-initialization combinations. The resulting number of Kronecker products is $4^K$, with $K$ being the number of edges cut. For general quantum circuits with $n$ quantum edges, this task faces an $O(n!)$ combinatorial search space. Section 4.1 addresses this problem with mixed-integer programming. Our work shows that with only a few cuts, many useful applications can be tractably mapped to NISQ devices currently available.

The second challenge is to scale the classical postprocessing. Large quantum circuits have exponentially increasing state space that quickly becomes intractable to even store the full-state probabilities. Section 4.3 addresses this problem with a dynamic definition algorithm to efficiently locate the "solution" states or sample the full output distribution for large quantum circuits beyond the current QC and classical simulation limit.

## 4 FRAMEWORK OVERVIEW

Figure 5 summarizes the key components of CutQC. Our framework is built on top of IBM's Qiskit [2] package in order to use IBM's quantum devices, but we note that the hybrid approach works with any gate-based quantum computing platforms. Given a quantum circuit specified as an input, the first step is to decide where to make cuts. We propose the first automatic scheme that uses mixed-integer programming to find optimal cuts for arbitrary quantum circuits. The backend for the MIP cut searcher is implemented in the Gurobi solver [18]. Small quantum devices then evaluate the different combinations of the subcircuits. Eventually, a parallel $C$ implementation of the **reconstructor** postprocesses the subcircuit outputs and reproduces the original full circuit outputs from the Kronecker products.

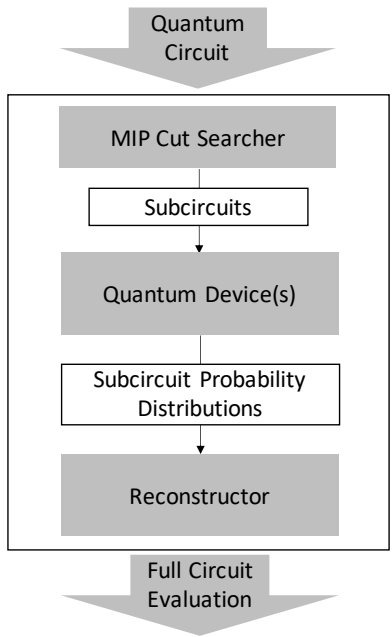

**Figure 5: Framework overview of CutQC. A mixed-integer programming (MIP) cut searcher automatically finds optimal cuts given an input quantum circuit. The small subcircuits resulting from the cuts are then evaluated by using quantum devices. The reconstructor then reproduces the probability distributions of the original circuit.**

## 4.1 MIP Cut Searcher

Unlike the manual example in Section 3.2, CutQC's cut searcher uses mixed-integer programming to automate the identification of cuts that require the least amount of classical postprocessing. Our problem instances are solved by the Gurobi mathematical optimization solver [18].

Without loss of generality, the framework assumes that the input quantum circuit is fully connected. That is, all qubits are connected via multiqubit gates either directly or indirectly through intermediate qubits. A quantum circuit that is not fully connected can be readily separated into fully connected subcircuits without cuts, and these do not need the classical postprocessing techniques to sew together. We focus on the more difficult general cases where cutting and reconstruction are needed.

*4.1.1 Model Parameters.* Besides an input quantum circuit, the MIP cut searcher requires the user to specify the maximum number of qubits allowed per subcircuit, $D$, equal to the size of the quantum devices available to the user. Another input is the maximum number of subcircuits allowed, $n_C$.

A quantum circuit can be modeled as a directed acyclic graph $G$. Quantum operations are always applied sequentially to the qubits, and neither classical nor quantum control dependencies are permitted under current hardware restrictions. The single-qubit gates are ignored during the cut-finding process, since they do not affect the connectivity of the quantum circuit. The multiqubit quantum gates are then modeled as the vertices $V = \{v_1, \ldots, v_{n_V}\}$, and the qubit wires are modeled as the edges $E = \{(e_a, e_b) : e_a \in V, e_b \in V\}$ in the graph. Choosing which edges to cut in order to split $G$ into subcircuits $C = \{c_1, \ldots, c_{n_C}\}$ can also be thought as clustering the vertices. The corresponding cuts can then obtained from the vertex clusters.

The MIP searcher uses a parameter $w$ associated with each vertex $v \in V$ that indicates the number of original input qubits directly connected to $v$. That is, $w_v \in \{0, 1, 2\}, \forall v \in V$. Note that $w$ depends only on the input quantum circuit. In this paper, $w_v$ can only take the values 0, 1, or 2 since we consider only circuits with gates involving at most two qubits. This approach is consistent with the native gates supported on current superconducting hardware.[1] Any gates involving more than two qubits can be decomposed into the native gate set before execution on quantum computers.

*4.1.2 Variables.* Inspired by constrained graph clustering algorithms [4], we define the following variables associated with the vertices and the edges.

$$y_{v,c} \equiv \begin{cases} 1 & \text{if vertex } v \text{ is in subcircuit } c \\ 0 & \text{otherwise} \end{cases}, \; \forall v \in V, \forall c \in C$$

$$x_{e,c} \equiv \begin{cases} 1 & \text{if edge } e \text{ is cut by subcircuit } c \\ 0 & \text{otherwise} \end{cases}, \; \forall e \in E, \forall c \in C$$

The number of qubits required to run a subcircuit is the sum of two parts, namely, the number of original input qubits and the number of initialization qubits induced by cutting (in Figure 4, $subcirc2_0$ is an example of an initialization qubit). The number of original input qubits, $\alpha_c$, in each subcircuit depends simply on the weight factors $w_v$ for the vertices in the subcircuit and is given by

$$\alpha_c \equiv \sum_{v \in V} w_v \times y_{v,c}, \forall c \in C. \tag{4}$$

A subcircuit requires initialization qubits when a downstream vertex $e_b$ is in the subcircuit for some edge $(e_a, e_b)$ that is cut. The number of initialization qubits, $\rho_c$, is hence

$$\rho_c \equiv \sum_{e : (e_a, e_b) \in E} x_{e,c} \times y_{e_b,c}, \forall c \in C. \tag{5}$$

A subcircuit requires measurement qubits when an upstream vertex $e_a$ is in the subcircuit for some edge $(e_a, e_b)$ that is cut. The number of measurement qubits, $O_c$, is

$$O_c \equiv \sum_{e : (e_a, e_b) \in E} x_{e,c} \times y_{e_a,c}, \forall c \in C. \tag{6}$$

Consequently, the number of qubits in a subcircuit that contributes to the final measurement of the original uncut circuit is

$$f_c \equiv \alpha_c + \rho_c - O_c, \forall c \in C. \tag{7}$$

*4.1.3 Constraints.* We next turn to constraints. We require that every vertex be assigned to exactly one subcircuit.

$$\sum_{c \in C} y_{v,c} = 1, \quad \forall v \in V \tag{8}$$

---

[1]Current superconducting architectures are limited to 1- and 2-qubit gates; other architectures (based on ion traps or neutral atoms) allow for multiqubit gates. The MIP cut searcher can easily be generalized to multiqubit gates.

We also require that the $d_c$ qubits in subcircuit $c$ be no larger than the input device size $D$.

$$d_c \equiv \alpha_c + \rho_c \leq D, \ \forall c \in C \tag{9}$$

To constrain the variable $x$, we note that an edge $e$ pointing from vertex $e_a$ to $e_b$ is cut by a subcircuit $c$ if and only if that subcircuit contains one and only one of these two vertices. An edge, if cut at all, is always cut by exactly two subcircuits. Thus, $x_{e,c} = 0$ indicates that either $e$ is not cut at all or that $e$ is cut somewhere else but just not in subcircuit $c$. The constraint on the variable $x$ is hence defined as

$$x_{e,c} = y_{e_a,c} \oplus y_{e_b,c}, \forall e = (e_a, e_b) \in E, c \in C. \tag{10}$$

This nonlinear constraint can be encoded by linear constraints:

$$\begin{aligned} x_{e,c} &\leq y_{e_a,c} + y_{e_b,c} \\ x_{e,c} &\geq y_{e_a,c} - y_{e_b,c} \\ x_{e,c} &\geq y_{e_b,c} - y_{e_a,c} \\ x_{e,c} &\leq 2 - y_{e_a,c} - y_{e_b,c}. \end{aligned} \tag{11}$$

For a given solution to this optimization problem, there are $n_C!$ possible relabelings with identical objective function values. Breaking all such symmetries can significantly decrease the time required to solve problem instances but can require introducing many auxiliary variables and constraints [35]. Nevertheless, our formulation breaks by forcing vertices with smaller indices to be in subcircuits with smaller indices. Specifically, we require vertex 1 to be in subcircuit 1, vertex 2 to be in subcircuit 1 or subcircuit 2, and so on. This requirement translates to the following constraint:

$$\sum_{j \geq k+1}^{n_C} y_{k,j} = 0, \quad k = 1, \ldots, n_C. \tag{12}$$

*4.1.4 Objective Function.* For efficiency and without loss of generality, we seek to minimize the classical postprocessing overhead required to reconstruct a circuit from its subcircuits. Therefore, the objective is set to be the number of floating-point multiplications involved in the *build* step.

The number of cuts made is given by

$$K = \frac{1}{2} \sum_{c \in C} \sum_{e \in E} x_{e,c}, \tag{13}$$

The objective function for the MIP cut searcher is hence the reconstruction time estimator:

$$L \equiv 4^K \sum_{c=2}^{n_C} \prod_{i=1}^{c} 2^{f_i}. \tag{14}$$

This cost objective accurately captures the bulk of the computation when we aim to build the full $2^n$ probabilities for an $n$-qubit uncut circuit, under the full definition CutQC mode (discussed in Section 4.2).

However, there is a prohibitive memory requirement for storing the $2^n$ probabilities as floating-point numbers when circuits get larger. Section 4.3 introduces a novel dynamic definition method to efficiently sample very large circuits with a much lower postprocessing overhead. Nevertheless, we chose to minimize Equation 14 during cut search as a positively correlated objective.

The overall MIP cut search problem is therefore

$$\begin{aligned} \text{minimize objective} \quad & L \text{ (Eq. 14)} \\ \text{s.t. constraints} \quad & \text{Eqs. (4)} - (13). \end{aligned} \tag{15}$$

## 4.2 Classical Postprocessing

We developed two types of classical postprocessing algorithms: a full-definition (FD) query and a dynamic-definition (DD) query algorithms. The difference in these methods lies in whether the entire $2^n$ full-state probability output of the uncut circuit is reconstructed.

FD query reconstructs the probability for every possible output state of the uncut circuit. To make the postprocessing more efficient, we developed three techniques: *greedy subcircuit order*, *early termination*, and *parallel processing*. The combination of these techniques improves the performance of the CutQC postprocessing toolchain. In addition, we used the kernel functions in the Basic Linear Algebra Subprograms (BLAS) through the Intel Math Kernel Library [51] to optimize the performance on CPUs.

The *greedy-subcircuit-order* technique exploits the fact that a large quantum circuit is (in general) cut into subcircuits of different sizes. The order of subcircuits in which the reconstructor computes the Kronecker products incurs different sizes of carryover vectors and affects the total number of floating-point multiplications. Our approach places the smallest subcircuits first in order to minimize the carryover in the size of the vectors. Since the reconstructor must eventually reproduce a probability vector with size equal to the Hilbert space of the uncut circuit, this technique may reduce the overhead by up to 50%.

The *early termination* technique exploits the fact that a Kronecker product term ends up being a vector of all zeros if any of its components contains only zeros. Such a Kronecker product term hence does not contribute to the full circuit output and can be ignored by the reconstructor. Experiments using classical simulation to produce the subcircuit outputs show that this situation happens surprisingly often. As a result, many Kronecker terms can be skipped by the reconstructor.

The *parallel processing* approach exploits the fact that the vector arithmetics have no data dependency at all and can hence be easily executed in parallel. Individual compute nodes read the subcircuit output data stored on disk in order to avoid the need for any internode communications. This approach allows our toolchain to scale almost perfectly with increasing numbers of compute nodes.

## 4.3 Dynamic Definition

Quantum circuits can be loosely categorized into two groups. The first group produces sparse output probabilities, where just a few "solution" states have very high probabilities and the "non-solution" states have zero probabilities. Most known quantum algorithms fall into this category, such as Grover search [17], the Bernstein–Vazirani algorithm [7], and the Quantum Fourier Transform [12]. This is where QC shows promise over classical computing by efficiently locating the "solution" states.

The second group of circuits produces dense output probabilities, where many states have nonzero probabilities. For this type of circuit, even with access to quantum computers large enough to execute the circuits directly, querying the FD probability output quickly becomes impossible. The reason is that (1) an exponentially increasing amount of memory is required to store the probabilities and (2) an exponentially increasing number of shots are required on a quantum computer before the probabilities converge. Fortunately, knowing the FD probabilities of all states simultaneously is usually not of interest. Instead, users are interested in the distribution itself. Examples include the 2-D random circuits from Google [9], which produce the Porter–Thomas distribution [38].

DD query allows us to find the "solution" states or sample dense probability distributions efficiently with very large quantum circuits, even when storing the full-state probability is not tractable. DD query produces a probability distribution that merges certain states into one bin and maintains the sum of their probabilities instead of the individual states within. Algorithm 1 presents the DD algorithm. In each recursion, DD runs the subcircuits and merges subcircuit states before postprocessing. The active qubits in each recursion determine the number of bins, and the merged qubits determine which states are merged in a single bin. Users can choose which qubits are active, that is, which qubits have their states actively explored. For those qubits, DD then recursively zooms in on their more prominent (i.e., higher probability) bins; this lets DD efficiently obtain fine-grained probabilities for the states within the bins.

---

**Algorithm 1:** Dynamic Definition

Initialize empty list *probability_bins*;
**for** *each DD recursion* **do**
  **if** *First recursion* **then**
    Choose a subset of qubits to label as *active*, max number determined by system memory;
  **else**
    Choose the bin from *probability_bins* with the largest sum of probability;
    Fix the *active* qubits in the bin according to the index of the bin; label as *zoomed*;
  Label the rest of the qubits as *merged*;
  Attribute the subcircuit shots, group shots with common *merged* qubits together;
  Reconstruct the $2^{\#active}$ probability output for the *active* qubits; append to *probability_bins*;

---

For sparse outputs, DD can recursively pinpoint the "solution" states and their probabilities. To do so, DD query follows a DFS-like search strategy to recursively choose the *qubit_state* with higher probabilities to zoom in on. By recursively locating the *active* qubits in their most probable *zoomed* states, "solution" states can be easily located after just a few recursions. For an $n$-qubit full circuit, the number of recursions needed is $O(n)$.

For dense outputs, DD can build a "blurred" probability landscape of the exact FD probability distribution, with the ability to arbitrarily "zoom in" on any region of the Hilbert space. To do so, DD query follows a BFS-like strategy to choose the *qubit_state* with higher probabilities to zoom in on. Users can decide the number of recursions and subset of states of interest to zoom in on. This is equivalent to efficient sampling of very large circuits on small quantum computers.

# 5 METHODOLOGY

## 5.1 Backends

We test our approach by running postprocessing and classical simulation benchmarks on a medium-size computing cluster using up to 16 compute nodes. Each node has an Intel Xeon CPU E5-2670 v3 at 2.30 GHz, with 256 GB allocated DDR4 memory [22]. We found 16 compute nodes to be sufficient to process the data generated by our tests (reported in the experiments section).

We first tested FD query for circuits up to 35 qubits, where storing the full-state probability is still tractable. We cut original circuits and mapped the resulting subcircuits to quantum computers of various sizes to demonstrate executing subcircuits larger than device sizes and study the postprocessing runtime. Because NISQ devices currently do not allow high-fidelity executions of circuits beyond just a few qubits, we executed the subcircuits with statevector simulation to demonstrate the effectiveness of our postprocessing techniques.

We tested DD query for circuits up to 100 qubits, significantly beyond the current classical and quantum limit. Because no backends are capable of producing accurate circuit executions on this scale, we used uniform distributions as the subcircuit output to study the runtime.

Since gate times of superconducting quantum computers are on the order of nanoseconds [3], we assume that quantum computer runtime is negligible in the comparisons. CutQC allows executing the subcircuits on many small quantum computers in parallel to further reduce the time spent on quantum computers. In addition, our experiments limit the MIP cut searcher to search for cuts that will divide an input circuit into at most 5 subcircuits with 10 cuts. The set of cuts with the smallest objective function value is then taken to be the optimal solution, with ties broken arbitrarily. MIP is able to find an optimal solution within minutes for all the experiments reported in this paper; its runtime therefore is also ignored.

## 5.2 Metric

For the runtime analysis, we allow CutQC to compute the independent Kronecker products during postprocessing for at least 10 minutes on each compute node, after which we scale the runtime according to the total number of Kronecker products required. For cases where < 10 minutes postprocessing is required, our benchmark reports the end-to-end wall time. We verified the scaled-up runtime against the end-to-end runtime for several medium-sized circuits to confirm the validity of our scaling approach. This approach is accurate because each Kronecker product requires the same amount of computation.

Besides the runtime analysis, we ran CutQC with IBM's 5-qubit Bogota device to compare the fidelity with directly executing the circuits on IBM's 20-qubit Johannesburg device. As NISQ devices improve, CutQC can be applied to larger devices to produce useful executions on larger scales. To quantify the noise behaviors, we used $\chi^2$ loss

$$\chi^2 = \sum_{i=0}^{2^n-1} \frac{(a_i - b_i)^2}{a_i + b_i}, \tag{16}$$

where $a_i$ are elements of circuit execution probability distributions (from Figure 2b, 2c) and $b_i$ are elements of the ground truth probability distributions (from Figure 2a). The smaller the $\chi^2$ is, the better the execution results are.

## 5.3 Benchmarks

We used the following circuits as benchmarks.

(1) *Supremacy*. This is a type of 2-D random circuit adapted from [9]. It is an example of circuits with dense probability output and was used by Google to demonstrate quantum advantage [3]. The circuit depth was 10 in our experiments. We verified that the rectangular shapes (such as 2*10) are much easier to be cut and require little postprocessing. We therefore focused only on the more difficult near-square shapes, with the two dimensions differing by up to 2 qubits (such as 4*5). Hence not all numbers of qubits are valid.

(2) *Approximate Quantum Fourier Transform* (*AQFT*). QFT [12] is a common subroutine in many quantum algorithms that promise speedup over classical algorithms. AQFT has been proposed to yield better results than QFT on NISQ devices [5].

(3) *Grover*. In comparison with classical algorithms this quantum Grover search algorithm offers polynomial speedup in unstructured database search [17]. We used the Qiskit [2] implementations of the Grover search that require $n-1$ ancillas; hence only odd numbers of qubits are valid. Additionally, Qiskit's oracle construction does not scale beyond 59-qubit Grover circuits.

(4) *Bernstein–Vazirani* (*BV*). This quantum algorithm solves the hidden string problem more efficiently than classical algorithms do [7].

(5) *Adder*. Adder is a quantum ripple-carry adder with one ancilla and linear depth [14]. It is an important subroutine in quantum arithmetic involving summing two quantum registers of the same width; hence only even numbers of qubits are valid.

(6) *Hardware efficient ansatz* (*HWEA*). HWEA is an example of near-term variational applications, a promising class of quantum algorithms on NISQ devices [30].

The benchmark circuits represent a general set of circuits for gate-based QC platforms and promising near-term applications.

## 6 EXPERIMENT RESULTS

### 6.1 Full Definition Query

The size of quantum devices serves as the baseline to demonstrate CutQC's ability to expand the size of quantum circuits. The Qiskit runtime of classically simulating the benchmark circuits serves as the baseline to demonstrate CutQC's ability to speed up quantum circuit evaluations.

The experiments in Figure 6 show the effect of different benchmarks, quantum circuit sizes, and quantum computer sizes on postprocessing runtime. We used 10-, 15-, 20-, and 25-qubit quantum computers and ran benchmark circuits larger than the devices in FD query using 16 compute nodes for postprocessing. We achieve an average of 60X to 8600X runtime speedup over classical simulation for our benchmarks.

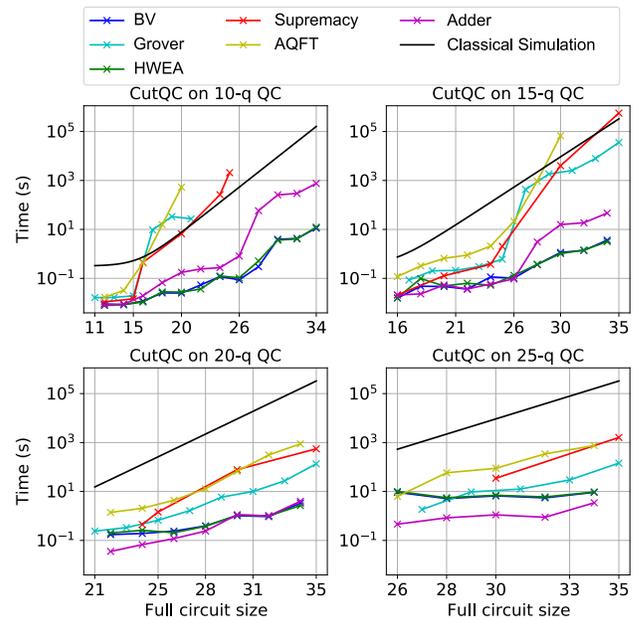

Figure 6: Use of CutQC to execute circuits mapped to 10-, 15-, 20-, and 25-qubit quantum computers in FD query. The horizontal axis shows the size of the quantum circuits; the vertical axis shows the postprocessing runtime in log scale. Experiments are done with all optimization techniques and on 16 compute nodes. BV and HWEA have similar runtimes, and the lines are not discernible in the 25-q plot. CutQC enables FD query almost always faster than classical simulations do. CutQC offers an average of 60X to 8600X runtime speedup over classical simulation alternatives for different benchmarks.

Some benchmarks cannot be mapped onto the quantum computers within 10 cuts and 5 subcircuits, Figure 6 thus has some of the benchmarks terminated early. *Supremacy*, *Grover*, and *Adder* face size limitations mentioned in Section 5, and we examine only an even number of qubits for *AQFT*, *BV*, and *HWEA*.

The type of benchmarks, quantum circuit sizes, and available quantum computer sizes are all important contributors to runtime. First, some benchmarks are harder to cut and require more postprocessing overhead. Specifically, *Supremacy*, *Grover*, and *AQFT* are more densely connected circuits and generally require more postprocessing. Second, larger quantum circuits generally require more postprocessing. The reason is that executing quantum circuits that significantly exceed the available quantum resources has to rely more on classical computing resources. In some cases, the classical postprocessing incurred outweighs any benefit from having quantum computers, and the resulting runtime is longer than classical simulation. Third, having larger quantum computers generally improves the runtime. However, having larger quantum computers faces diminishing returns. The postprocessing overhead eventually plateaus when the quantum computer is large enough to support an efficient partitioning of the circuit. For example, the 5*7 *Supremacy* circuit is cut into 2 subcircuits with 5 cuts on both 20- and 25-qubit computers and has similar runtime.

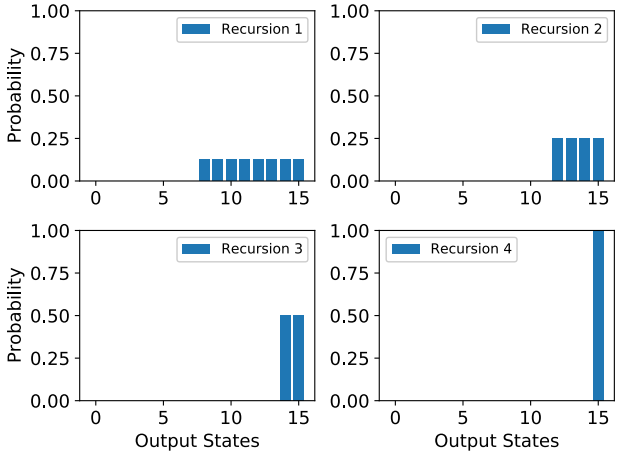

**Figure 7: Use of CutQC to execute a 4-qubit *BV* circuit on 3-qubit quantum computers in DD query. During each recursion, we plot the probability of every state in a merged bin as the average of the sum of probabilities for that bin. With recursive zoom-in, recursion 4 shows that the sum of probability for the output state |1111⟩ is 1; that is, it is the solution state.**

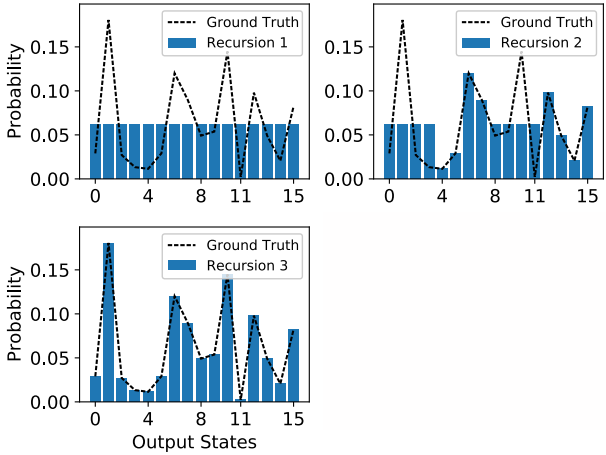

**Figure 8: Use of CutQC to execute a 4-qubit *supremacy* circuit on 3-qubit quantum computers in DD query. By recursively zooming in and improving the definition for bins with higher probabilities, CutQC allows building a better approximation to the ground truth distribution.**

## 6.2 Dynamic Definition Query

We used DD to efficiently sample quantum circuits of which the full Hilbert space is too large to even store. We first used 4-qubit *BV* and *Supremacy* circuits to illustrate the DD process of locating the solution state and sampling a target probability distribution.

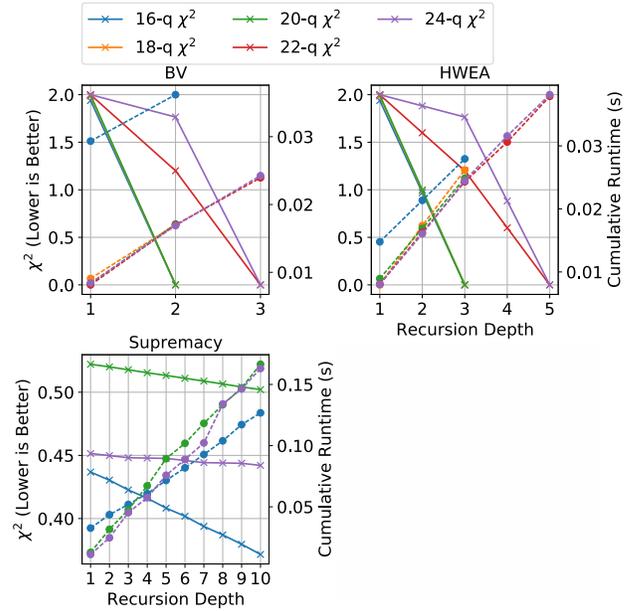

**Figure 9: Use of CutQC to execute 16- to 24-qubit benchmark circuits on 15-qubit quantum computers in DD query for a maximum of 10 recursions or until $\chi^2 = 0$. Maximum system memory is set to 10-qubit. Solid lines (left axis) show the decreasing $\chi^2$ as DD obtains more fine-grained information about the full distribution. Dotted lines (right axis) show the increasing cumulative runtime of all recursions. In contrast, classical simulation of 24-qubit circuits takes about 130 seconds on the same classical hardware.**

Figure 7 shows results after cutting and executing a 4-qubit *BV* on 3-qubit quantum computers. We set the number of *active* qubits during each recursion to 1; hence 4 recursions are required to locate the solution state. Recursion 1 *merges* qubits 2, 3, and 4 and shows that the sum of probabilities for output states |0000⟩ to |0111⟩ is 0 and for |1000⟩ to |1111⟩ is 1. Recursion 2 then holds qubit 1 in state |1⟩ and *merges* qubits 3 and 4 to show that the sum of probabilities for output states |1000⟩ to |1011⟩ is 0 and for |1100⟩ to |1111⟩ is 1. Eventually, recursion 4 successfully locates the solution state to be |1111⟩. Each recursion stores and computes vectors only of length $2^1$, instead of $2^4$. DD on larger circuits works similarly.

For circuits with dense output, DD chooses to zoom in and increase the definition for bins with higher probabilities, in order to reconstruct a blurred probability landscape. Figure 8 shows results after cutting and executing a 4-qubit *Supremacy* circuit on 3-qubit quantum computers. Each recursion zooms in on a bin of states with the highest probability, improving its definition. More recursions hence allow a closer reconstruction of the ground truth probability landscape.

Figure 9 shows the evolution of $\chi^2$ via DD for several medium sized benchmark circuits. *BV* has exactly one solution state and hence requires just a few recursions to locate it. *HWEA* has two solution states that are maximally entangled. DD recursively improves the definition of the more prominent bins and is also able to locate the solution states in just a few recursions. *Supremacy* has dense output, its $\chi^2$ decreases as DD runs more recursions, without ever storing the full distribution. Figure 9 evaluates up to 24 qubits and 10 recursions because the computation of $\chi^2$ takes a long time for even medium-sized quantum circuits. It is noteworthy that the per-recursion and cumulative runtime of CutQC is negligible compared with the purely classical simulation runtime on the same classical hardware.

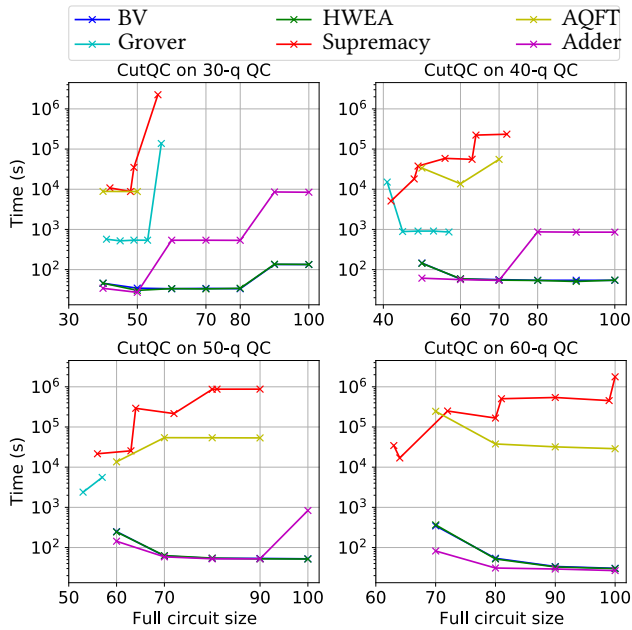

**Figure 10: Use of CutQC to execute circuits mapped to 30-, 40-, 50-, and 60-qubit quantum devices in DD query. The vertical axis shows the postprocessing runtime of 1 DD recursion with a definition of $2^{35}$ bins.**

NISQ devices will gradually improve in fidelity and sizes to allow evaluating subcircuits beyond the classical simulation limit. CutQC then will allow the use of those NISQ devices to efficiently evaluate even larger quantum circuits. We cut and executed circuits of up to *100* qubits and used DD query to sample their blurred probability landscape with a definition of $2^{35}$ bins in one recursion. Figure 10 shows the runtime of using 30-, 40-, 50-, and 60-qubit quantum computers. Larger quantum computers allow the execution of larger quantum circuits and faster runtime. Therefore, DD offers a way to efficiently sample large circuits, with the ability to arbitrarily increase the definition of any subregions of interest by doing more recursions. This is all without the need for either a large quantum computer or vast classical computing resources.

### 6.3 Real QC Runs

To study the effect of device noise on our toolchain, we ran experiments on IBM's real quantum devices. Figure 11 compares the circuit output obtained from (a) directly executing circuits on the state-of-the-art 20-qubit Johannesburg device and (b) executing circuits with more than 5 qubits on the 5-qubit Bogota device with CutQC. We show that CutQC evaluation with small quantum computers produces a lower $\chi^2$ loss and hence outperforms QC evaluation with large quantum computers. CutQC reduces $\chi^2$ loss by nearly 60% in the best cases. The experiments stop at 12 qubits because QC evaluation beyond this point succumbs to the effects of noise and fails to produce meaningful output. Among the benchmarks, only the AQFT circuits experienced a negative reduction. This is because AQFT compiled for the current NISQ devices is much deeper than the other benchmarks. Therefore both QC and CutQC on AQFT have accuracy too low for meaningful comparisons. As NISQ devices improve in noise and connectivity, we expect AQFT to improve.

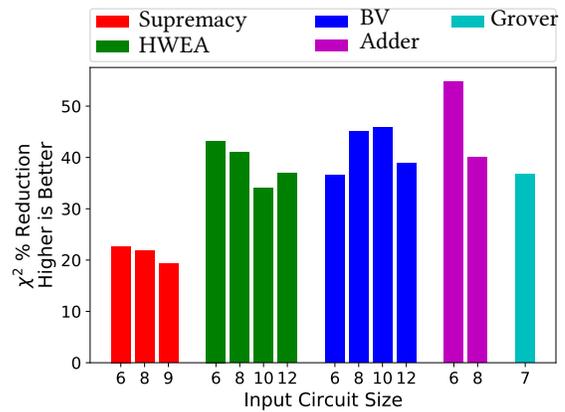

**Figure 11: Comparison of the 20-qubit Johannesburg quantum computer versus the 5-qubit Bogota device with CutQC. For each benchmark we find the ideal output distribution via statevector simulation. We then use this ideal distribution to compute the $\chi^2$ metric for two execution modes: QC evaluation on the Johannesburg device ($\chi_J^2$) and CutQC evaluation utilizing the Bogota device ($\chi_B^2$). The reported $\chi^2$ percentage reduction is computed as $100 * (\chi_J^2 - \chi_B^2)/\chi_J^2$. A distribution that is close to ideal will have a small $\chi^2$ value, and therefore a positive $\chi^2$ percentage reduction indicates improved performance. Only the AQFT workloads experience a negative reduction and are omitted. CutQC achieves an average of 21% to 47% $\chi^2$ reduction for different benchmarks.**

Despite requiring more subcircuits and readout, CutQC evaluates circuits with better fidelity. The main reason for such improvements is that CutQC runs subcircuits that are both smaller and shallower than the uncut circuit run by the QC mode. Furthermore, CutQC substitutes the noisy quantum entanglement across subcircuits by noise-free classical postprocessing. Quantum compilers also are suboptimal [45] and may affect QC more than they do CutQC, because the logical circuits they need to compile are more complicated in the QC mode.

Not only does CutQC need smaller quantum computers, it also produces better outputs. Therefore, combined with CutQC, building small but reliable quantum computers becomes much more useful than merely increasing qubit counts at the cost of degrading fidelity.

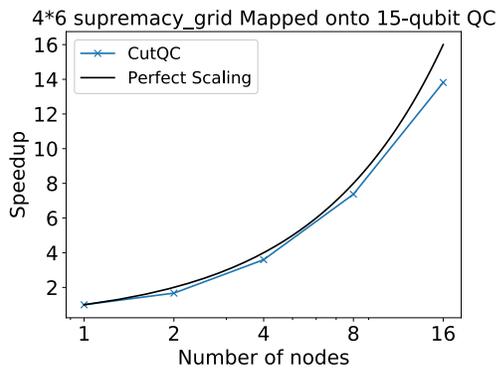

**Figure 12: Postprocessing a 4\*6 supremacy_grid circuit mapped to a 15-qubit IBM Melbourne device. Four cuts incur $4^4 = 256$ Kronecker products. The 16-node postprocessing requires roughly 0.5 seconds, a 260X speedup over classical simulation alternative. The blue line shows the postprocessing runtime speedup compared with that of a single node. The black line shows perfect scaling as a reference. Runtime scales well with the number of compute nodes.**

## 6.4 Discussion – Comparison with Classical Simulations

Other classical simulation techniques partition qubits, decompose 2-qubit gates across the partitions, simulate each partition classically, and employ Feynman path simulation [10, 28]. These methods have key differences from CutQC: (1) they cannot run on NISQ devices because they require complex amplitudes of the states, which are not available from NISQ devices; and (2) they cut 2-qubit gates across qubit partitions, instead of quantum edges among gates. Because of the differences in methods, implementations, and experimental setup, direct comparisons are difficult. However, Feynman path simulations do not scale well past subcircuits beyond the classical simulation limit, while CutQC easily scales as NISQ devices become larger and more reliable, as Figure 10 demonstrates.

As one example, [28] only simulates $2^{20}$ states with < 1% fidelity for up to 56-qubit input quantum circuit. In contrast, CutQC processes $2^{35}$ bins in Figure 10 for up to 100 qubits with perfect postprocessing fidelity. The overall fidelity of CutQC is limited by NISQ noise, but this will improve as NISQ devices improve, and these improvements will not increase CutQC's runtime.

In addition, other classical simulation results make use of supercomputers to perform quantum circuit evaluations in parallel. Directly comparing with such results is challenging because many simulations often require supercomputers with hundreds to thousands of compute nodes [19, 50, 52], millions of core-hours [49], and a prohibitive amount of memory [36]. Furthermore, many simulate only a small subset of output states for large quantum circuits, called partial state simulation [11, 50]. Most of these approaches do not scale. CutQC offers advantages in runtime, resources requirement, and the ability to sample the full output distribution.

First, CutQC requires no internode communication and hence has nearly perfect multinode scalability. For example, we cut and execute a 4\*6 *Supremacy* circuit on the 15-qubit Melbourne quantum computer. Figure 12 shows the postprocessing runtime speedup as the number of parallel nodes increases. The 16-node postprocessing has a 14X speedup over 1 node. CutQC does not require many compute nodes when the postprocessing incurs only a few Kronecker products. When more cuts are required, we expect the runtime to scale well with more compute nodes. Our scaling studies with 1–16 nodes indicate that CutQC can be easily ported to CPU-based supercomputing platforms to scale to thousands of compute nodes.

Second, the DD query algorithm efficiently samples the full output distribution with good scalability. Partial state simulation produces the probability only for very few output states, representing an infinitesimal region of the entire Hilbert space, whereas the DD query efficiently samples the entire Hilbert space with scalable runtime (Figure 10).

## 7 RELATED WORK

Many quantum compilation techniques have been developed with the goal of improving the performance of NISQ devices. Using real-time device calibration data to map logical qubits to physical qubits [32, 48], efficiently scheduling operations to reduce quantum gate counts [20, 44], and repeating circuit executions to mitigate error [23, 46, 47] are among the more recently developed techniques. However, these focus on improving a purely quantum computing approach and are intrinsically limited by the size and fidelity of NISQ devices. Specifically, our experiments used the noise adaptive compiler [32] in both CutQC and QC evaluations. The improved fidelity we demonstrate is in addition to that given by the compiler. Furthermore, previous compilers do not allow executions of circuits beyond quantum computer sizes at all. Our approach can work in concert with any compilers to execute circuits both larger in size and better in fidelity.

Theoretical physics approaches have considered trading classical and quantum computational resources. These approaches, however, use simple partitioning of qubits [10] or involve exponentially higher postprocessing [37]. Several works manually separate small toy circuits with convenient structures as proof-of-concept demonstrations [6, 53]. Our approach is more flexible, has exponentially lower overhead, automatically selects cut positions, works with circuits of arbitrary structures, and is the first end-to-end scalable toolchain.

Previous works on classical simulation require massive computing resources [19, 36, 49, 50, 52], or only simulate very few output states with low fidelity [11, 28, 50].

## 8 CONCLUSION

This paper demonstrates how to leverage both quantum and classical computing platforms together to execute quantum algorithms of up to 100 qubits while simultaneously improving the fidelity of the output. Our results are significantly beyond the reach of current quantum or classical methods alone, and our work pioneers pathways for scalable quantum computing. Even as NISQ machines scale to larger sizes and as fault-tolerant QC emerges, CutQC's techniques for automatically cutting and efficiently reconstructing quantum circuit executions offer a practical strategy for hybrid quantum/classical advantage in QC applications.

## ACKNOWLEDGMENTS

We thank Yuri Alexeev and François-Marie Le Régent for helpful discussions and Zain Saleem for valuable feedback.

Funding acknowledgments: W. T., and T. T. are funded by EPiQC, an NSF Expedition in Computing, under grant CCF-1730082. The work of W. T., T. T., and M. S. is also supported by Laboratory Directed Research and Development (LDRD) funding from Argonne National Laboratory, provided by the Director, Office of Science, of the U.S. Department of Energy under contract DE-AC02-06CH11357. J. L. and M. S. are also supported by the U.S. Department of Energy, Office of Science, Office of Advanced Scientific Computing Research, Accelerated Research for Quantum Computing program. This research used resources of the Oak Ridge Leadership Computing Facility, which is a DOE Office of Science User Facility supported under Contract DE-AC05-00OR22725.

## A   ARTIFACT APPENDIX

### A.1   Abstract

Our artifact provides the source codes for the end-to-end CutQC toolflow. We also provide the benchmarking codes for several sample runtime and fidelity experiments. The HPC parallel version of the code is not provided, as different HPC platforms require very different setups.

### A.2   Artifact Checklist

- **Algorithm:** Mixed Integer Programming, Dynamic Definition, and Quantum Mechanics.
- **Compilation:** Intel icc (ICC) 19.1.0.166 20191121.
- **Hardware:** IBM superconducting quantum computers.
- **Run-time state:** Runtime is sensitive to CPU usage.
- **Metrics:** Wall clock runtime, $\chi^2$ defined by Equation 16.
- **Output:** Runtime and fidelity are printed by running the Python script provided.
- **Experiments:** We use CutQC package to measure the runtime and fidelity of several quantum circuits.
- **How much disk space required (approximately)?:** About 4GB.
- **How much time is needed to prepare workflow (approximately)?:** Under one hour.
- **How much time is needed to complete experiments (approximately)?:** About 20 minutes for the provided runtime benchmarks in the Python script. The fidelity script runtime largely depends on the real-time queue on IBMQ devices. Large customized quantum circuits take longer.
- **Archived (provide DOI)?:** https://doi.org/10.5281/zenodo.4329804.

### A.3   Description

*A.3.1  How to Access.* The DOI of our artifact is available at: https://doi.org/10.5281/zenodo.4329804.

*A.3.2  Hardware Dependencies.* The runtime experiments used classical computing hardware specified in Section 5. Different hardware may produce different runtimes, although the relative speedup should be similar.

The fidelity experiments require an active IBMQ account to access the IBM quantum computers.

*A.3.3  Software Dependencies.* Python 3.7, Intel icc (ICC) 19.1.0.166, Qiskit 0.23, Gurobi 9.0+.

### A.4   Installation

Users should refer to the *README.md* in the artifact directory to install the required software tools.

### A.5   Experiment Workflow

To run the runtime benchmark, run the Python script named "runtime_test.py". The script should take about 20 minutes to finish. To run the fidelity benchmark, run the Python script named "fidelity_test.py". The runtime largely depends on the real-time queue on IBMQ devices.

### A.6   Evaluation and Expected Results

Speedup over the Qiskit classical simulation is printed to terminal after running the "runtime_test.py" Python script. The provided script runs experiments on a 15-q QC. $\chi^2$ is printed after running the "fidelity_test.py" Python script. Expected results are reported in Section 6.

### A.7   Experiment Customization

User can adjust the relevant parameters in the provided Python scripts to run other runtime experiments. The adjustable parameters include size of QC, size of quantum circuits, type of circuits, number of parallel threads, max system memory, and IBM devices.

### A.8   Notes

Due to the different setup required to run on different HPC platforms, the multi-node parallel version of CutQC is not provided in the artifact. Instead, the artifact runs on single node with parallel threads. As a result, the speedup obtained from the artifact may be smaller than reported in the paper.

Furthermore, since the 20-qubit IBM *Johannesburg* device used in the paper has retired, we recommend using the 20-qubit *Boeblingen* device. Smaller experiments can also run on the publicly available 15-qubit *IBMQ_16_Melbourne* and the 5-qubit *Vigo* devices, although the fidelity results may vary.

### A.9   Methodology

Submission, reviewing and badging methodology:

- https://www.acm.org/publications/policies/artifact-review-badging
- http://cTuning.org/ae/submission-20201122.html
- http://cTuning.org/ae/reviewing-20201122.html